\documentclass[twocolumn,floatfix,prb,showpacs]{revtex4}
\usepackage{graphicx}
\usepackage{amsmath}
\usepackage{bm}
\usepackage{hyperref}
\usepackage{soul}
\usepackage{ulem}
\usepackage[dvips]{color}

\begin{document}
    \title{Unconventional superconducting states of interlayer pairing in bilayer and trilayer graphene}

\author{Mir Vahid Hosseini and Malek Zareyan}

\affiliation{Department of Physics, Institute for Advanced Studies
in Basic Sciences (IASBS), Zanjan 45137-66731, Iran}

\begin{abstract}
We develop a theory for interlayer pairing of chiral electrons in graphene materials which results in an unconventional superconducting (S) state with  s-wave spin-triplet order parameter.  In a pure bilayer graphene, this superconductivity exhibits a gapless property with an exotic effect of temperature-induced condensation causing an increase of the pairing amplitude (PA) with increasing temperature. We find that a finite doping opens a gap in the excitation spectrum and weakens this anomalous temperature-dependence. We further explore the possibility of realizing variety of pairing patterns with different topologies of the Fermi surface, by tuning the difference in the doping of the two layers. In trillayer graphene, the interlayer superconductivity is characterized by a two components order parameter which can be used to define two distinct phases in which only one of the components is non vanishing.  For ABA stacking the stable state is determined by a competition between these two phases. By varying the relative amplitude of the corresponding coupling strenghes, a first order phase transition can occur between these two phases.  For ABC stacking, we find that the two phases coexist with a possibility of a similar phase transition which turns out to be second order.
\end{abstract}

\maketitle

\section{introduction}
The microscopic theory of superconductivity developed by Bardeen, Cooper and Schrieffer\cite{BCS}(BCS) is based on the pairing between electrons from opposite spin-subbands with the same Fermi surfaces.
Generalization of the BCS pairing to composite systems of two (or more) types of fermions with different Fermi surfaces\cite{BCSWil} has attracted great interest due to its appearance in various areas of physics including pairing of ultracold Fermi atoms\cite{ColdAtoms,ColdAtomsExper}, color superconductivity in dense quark matter\cite{color1}, and neutron-proton pairing in nuclear systems \cite{NuclMatt}. Despite the robustness of conventional metallic superconductivity\cite{Tinkham} against small mismatches between the paired Fermi surfaces, several unconventional pairing phases have been predicted to realize in the presence of larger mismatches. The most well known cases include the phase of breached pairing (BP) referring to a state of separated normal (N) and S phases in the momentum space \cite{BreachedPairing}, pairing between deformed Fermi surfaces with zero total momentum of Cooper pairs  \cite{deformedFermi} and inhomogeneous S phase of Larkin-Ovchinnikov-Fulde-Ferrel (LOFF) \cite{LOFF}. These unconventional phases are resulted from different topologies of the Fermi surface.

The search for new exotic S state is the frontier subject of low temperature physics, because it sheds light on the pairing symmetry and may lead to understanding pairing mechanism of superconductivity in high-T$_c$~\cite{highTc} superconductors and other newly discovered S materials~\cite{Irons}. The recent discovery of graphene ~\cite{GeimScie,GeimNatu}, the two dimensional solid of carbon atoms with honeycomb lattice structure, and its associated bilayer and trilayer structures are expected to provide still a new opportunity for realizing unconventional pairing states. Graphene has a specific zero-gap electronic band structure in which the charge carriers behave like 2D massless Dirac fermions with a pseudo-relativistic chiral property. In addition to regular spin, electrons in graphene possess two additional quantum degrees of freedom, the so called pseudospin and valley. These features together with the fact that in graphene the carrier type, [electron-like (n) or hole-like (p)] and its density can be tuned conveniently, make these carbon-based material exceptional for realizing unconventional superconductivity.  The unusual features of superconductivity has been already predicted in monolayer graphene\cite{SuperTheorMonLayer}, where the pairing of electrons with opposite sublattice pseudo-spin leads to appearance of an unusual spin-singlet p+ip-wave S phase with no gap in its excitation spectrum\cite{SuperTheorMonLayerCastro}.  An intrinsic superconductivity, with plasmon or phonon mediated pairing interactions, can be realized in graphene coated with a metal \cite{SuperTheorMonLayerCastro,grapheneSuperTher}. For a coated bilayer and trilayer graphene, the formation of the S state is expected to be closely similar to that of the graphite intercalated with alkaline metals, for which a critical temperature up to 11.5 K has been reported \cite{SuperIntercalacte}. Furthermore, progress has already been made in proximity induced superconductivity by fabrication of transparent contacts between a graphene monolayer and a superconductor (see for instance, Refs. \cite{grapheneSuperExper}).
In a very recent study~\cite{HosseiniSpinTriplet}, we have explored the exotic nature of interlayer superconductivity in pure bilayer graphene.  It has been explained that how the interplay between the interlayer pairing of electrons with the same sublattice chirality and the asymmetric arrangement of the sublattices of the two layers result in a gapless superconductivity with an unusual s-wave spin-triplet symmetry of the order parameter and anomalous thermodynamic properties. We have obtained that the interlayer pairing allows for the possibility of a temperature-induced condensation causing an increase of the PA with increasing temperature, and an entropy of the stable S state which can be higher than its value in N state.
Motivated by these findings for interlayer superconductivity of pure bilayer graphene, we study effect of a finite doping.  Furthermore, considering the recent interest in the properties of trilayer graphene \cite{TrilayerStaking,Trilayerband}, we extend our study to the interlayer pairing in trilayer graphene. For bilayer graphene, we find that the doping opens a gap in the excitation spectrum which, in turn, weakens the temperature-induced condensation such that at high levels of the doping the temperature dependence of PA became similar to that of the conventional BCS gap\cite{Tinkham}. We present phase diagrams of interlayer superconductivity for both symmetrical and asymmetrical dopings with the two layers having the same and different levels of the doping, respectively.
\par
For trilayer graphene, we examine  the interlayer pairing in two different types of ABA and ABC stacking in pure case and find an strong dependence on the type of the stacking. We show that the asymmetric ABC stacking can support stronger pairing gap than the symmetric ABA stacking. For ABA stacking,  there is a competition between two phases  of interlayer superconductivity in  which the pairing is realized between chiral electrons of the middle layer with those in only one of the other two layers. This depends on the relative values of the two corresponding coupling strengths.  We explore the possibility of a  phase transition between the two phases by varying the coupling strengths which turns out to be first order for ABA stacking but second order for ABC stacking.

The paper is organized as follows. In the next section, we discuss the theoretical modeling of the interlayer superconductivity in graphene materials and study the resulted band structure with different topologies of Fermi surfaces. In Sec. \ref{bilayer}, we discuss the numerical results for the phase diagram and order parameter as function of the various parameters of both symmetrically and asymmetrically doped bilayer graphene. Sec. \ref{trilayer} is devoted to the case of interlayer superconductivity for trilayer graphene. Finally, in Sec. \ref{con}, we present our summary and conclusions.

\section{Theory} \label{s1}
In order to study interlayer superconductivity, we consider a model based on graphene materials. We introduce the theory of interlayer superconductivity to the case of bilayer graphene and then this theory shall be developed to the case of trilayer graphene.
Bilayer graphene is composed of two coupled graphene monolayers with $A_1$ and $B_1$ triangular sublattices in the top layer and $A_2$ and $B_2$ triangular sublattices in the bottom layer according to Bernal stacking in which every $A_1$ site of the top layer lies directly above a $B_2$ site of the bottom layer. In the absence of superconductivity,
the following Hamiltonian can be used to describe $\pi$ electrons of the bilayer graphene \cite{BilayerTB},

\begin{eqnarray}
\label{HN} H^{AB}_0=\sum^2_{\textit{l}=1}H^{mono}_{\textit{l}}+H^{AB}_{\perp},
\end{eqnarray}
with the monolayer Hamiltonian,
\begin{eqnarray}
\label{mono} H^{mono}_{\textit{l}}=- \sum_{\sigma ,i }\mu_{\textit{l}}n_{\textit{l} ,i,\sigma}
-t\sum_{\sigma ,\langle i,j\rangle }(a_{\textit{l}
,i,\sigma}^{\dagger}b_{\textit{l} ,j,\sigma}+H.c.),
\end{eqnarray}
 and the nearest neighbor interlayer hopping Hamiltonian,
\begin{eqnarray}
\label{HinterHop} H^{AB}_{\perp}=-t_{\perp}\sum_{\sigma, i}(a_{1,i,\sigma}^{\dagger}b_{2,i,\sigma}+H.c.),
\end{eqnarray}
where $a_{\textit{l} ,i,\sigma}$ ($b_{\textit{l} ,i,\sigma}$) and $a^{\dagger}_{\textit{l} ,i,\sigma}$ ($b^{\dagger}_{\textit{l} ,i,\sigma}$)
are the annihilation and the creation operators of an electron in the $i$th unit cell in the sublattice A(B) and the layer $\textit{l}$.
$\sigma=\pm$ denotes the spin state of the electron; $n_{\textit{l} ,i,\sigma}$ is the corresponding on-site particle density operator. The intralayer nearest neighbor hopping energy $t\approx 3 eV$ determines the Fermi velocity in graphene as $v_F \approx10^6 m/s$, and $t_\perp \approx0.4eV$ ($t_\perp/t \approx 0.13$) is the dominant interlayer hopping energy between the nearest neighbors $A_1$ and $B_2$; the chemical potential $\mu_{\textit{l}}$ can be controlled by a top- and bottom-gate electrodes independently.

By introducing attractive interaction between the electrons of the sublattice $A_1$ and $B_2$ through the following potential,

\begin{figure}[!tb]
\begin{center}
\includegraphics[width=8cm]{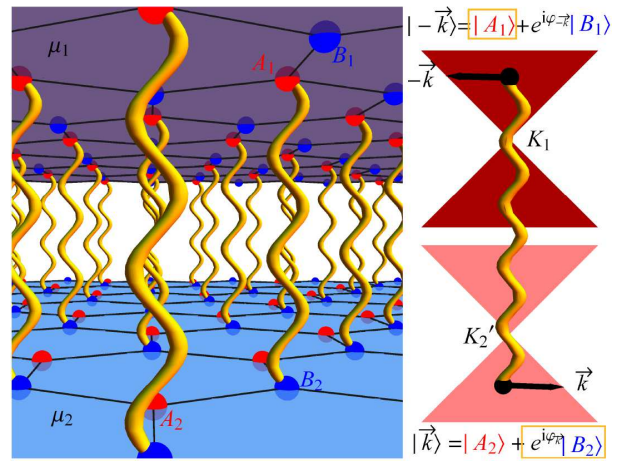}
\caption{(color online)
Left panel: Lattice structure of bilayer graphene. Interlayer S correlations in the real space are shown by wavy lines which couple electrons of different layers with opposite pseudospin degree of freedom. Right panel: Interlayer superconductivity in momentum space is shown by coupling of two time-reversed momentum states  $|-k\rangle$ and  $|k\rangle$ which are located in valleys $K_{1}$ and $K^{\prime}_{2}$ and belong to the different layers. Since, interlayer pairing is antisymmetric with respect to the pseudospin degree of freedom, pairing between components $|A_{1}\rangle$ and $|B_{2}\rangle$ is possible.
}
\label{pairing}
\end{center}
\end{figure}

\begin{eqnarray}
V^{AB}_{\perp}=-g_{\perp1} \sum_{\sigma,\sigma^{\prime},i}a_{1, i, \sigma}^{\dagger} a_{1, i, \sigma}b_{2, i, \sigma^{\prime}}^{\dagger}b_{2, i, \sigma^{\prime}},
\label{Vp}
\end{eqnarray}
the interlayer superconductivity can be produced. Here, $g_{\perp1}$ is the S coupling energy \cite{BaskaranCC,grapheneSuperTher}.
The interaction potential (\ref{Vp}) has on-site local property in the 2D plane of the bilayer graphene. Therefore, the wave function of two-body problem associated with the two coupled electrons and the corresponding pairing potential are not extended in the bilayer plane. This results in the introduction of isotropic s-wave symmetry for the orbital part of the order parameter. Because the coupling occurs between the two electrons with the opposite pseudospin degree of freedom, Pauli exclusion principle imposes a pairing in the spin-triplet channel. Therefore, the total wave function, including the product of orbital, spin and pseudospin parts will be antisymmetric under the exchange of electrons. The following s-wave spin-triplet order parameter can be used to decouple the total Hamiltonian of the bilayer graphene ($H^{AB}=H^{AB}_0+V^{AB}_{\perp}$),
\begin{eqnarray}
\Delta_{i,\perp1}=-g_{\perp1}\langle a_{1, i, \downarrow}b_{2, i, \uparrow}+a_{1, i, \uparrow}b_{2, i, \downarrow} \rangle.
\label{Order}
\end{eqnarray}
In the lattice space, as is shown in the left panel of Fig.~\ref{pairing}, $\Delta_{i,\perp1}$ describes the interlayer pairing of the two electrons with the sublattices $A_1$-$B_2$, and the spins $\uparrow$-$\downarrow$ (and $\downarrow$-$\uparrow$). In the momentum space, for the case of $t_{\perp}=0$ that the electronic band structures of the two monolayers are not affected by each other, there is a simple description for binding of the two time-reversed electronic states due to presence of attractive interaction as we discuss in the following.

At low energies, the chiral momentum states $|k\rangle$ located in Dirac cone-shape band structures at the two inequivalent valleys are coherent superposition of sublattice pseudospin states $|A\rangle$ and $|B\rangle$,
\begin{eqnarray}
|k\rangle=|A\rangle+exp(i\varphi_{k})|B\rangle,
\end{eqnarray}
where $\varphi_k$ is the angel of the momentum direction. Interlayer superconductivity induces partial pairing between the time-reversed momentum states $|k\rangle$ and $|-k\rangle$ located in the Dirac cones of different layers such that the coupling only takes place between $|A_1\rangle$ and $|B_2\rangle$ pseudospin parts of the electrons wave function in layers 1 and 2, respectively. Since the pairing of the time reversed chiral momentum states is partial, the interlayer superconductivity (\ref{Order}) gives raise even parity of the order parameter in the frequency space\cite{Oddfrequency}.

\begin{figure}[!tb]
\begin{center}
\includegraphics[width=6.cm]{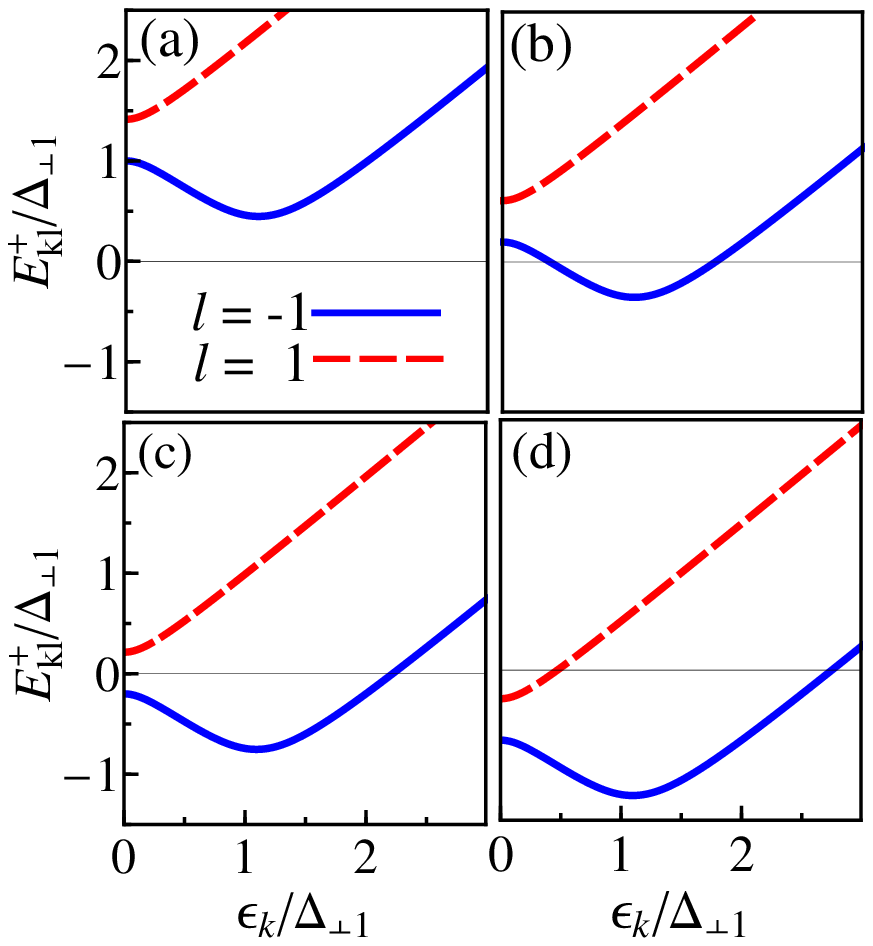}
\caption{(color online) The plot of quasi-particle excitation spectrum $E^{+}_{k \textit{l}}$ versus $\epsilon_k$ in S state. The dispersions correspond to the regimes of (a) BCS pairing with no effective Fermi surface for both branches $\textit{l}=\pm1$, (b) Sarma pairing with two effective Fermi surfaces for the branch $\textit{l}=-1$ but no Fermi surface for the other branch $\textit{l}=1$, (c) pairing with single Fermi surface in $\textit{l}=-1$  but no Fermi surface for the other branch $\textit{l}=1$, and (d) pn pairing with single Fermi surface in each branch $\textit{l}=\pm1$. }
\label{fig1}
\end{center}
\end{figure}

After decoupling of the interacting part of $H^{AB}$ by the order parameter (\ref{Order}), one can diagonalize Hamiltonian in 2D momentum space. The general expression of the spectrum is large, but for $t_{\perp}=0$, we obtain a simple expression for the spectrum $E_{\bf k \textit{l}}^{\gamma}$ which is,

\begin{eqnarray}
\alpha E_{\bf k  \textit{l}}^{\gamma}&=&\alpha [\sqrt{\frac{\Delta_{\perp1}^2+2(\mu^2+\epsilon_{\bf k}^2)+\textit{l} A}{2}}-\gamma h],
\label{eigen1}
\end{eqnarray}
\begin{eqnarray}
A&=&\sqrt{\Delta_{\perp1}^4+4\epsilon_{\bf k}^2(\Delta_{\perp1}^2+4\mu^2)},
\label{eigen2}
\end{eqnarray}
where $\textit{l},\gamma,\alpha=\pm$ indicate different branches of the spectrum. Here $\epsilon_{\bf k}=(3/2) t|\bf{k}|$, $\mu=(\mu_1+\mu_2)/2$ and $h=(\mu_1-\mu_2)/2$ are free band dispersion, mean doping and doping difference, respectively. Without loss of generality, we assume $\mu_1\geq\mu_2$. Branch $E_{\bf k  \textit{l}}^{-}$ for any value of parameters h, $\mu$, $\Delta_{\perp1}$, and $\textit{l}$ does not cross Fermi surface such that for breaking a paired state and sending quasiparticles to this state a finite energy is needed. This situation resembles metallic superconductor. Nevertheless, $E_{\bf k  \textit{l}}^{+}$ can cross Fermi energy through one point for $\textit{l}=1$ and either one or two points for $\textit{l}=-1$. As a result, it generates gapless excitation modes. In Fig. \ref{fig1} the quasiparticle excitation spectrum is shown. There are four possible features for $E_{\bf k  \textit{l}}^{+}$. Each feature with its own topology of Fermi surface can be accessible by tuning the value of h. The boundaries between four ranges of h associated with different topologies of the Fermi surfaces can be determined by analyzing the roots of $E^{+}_{k \textit{l}}=0$ which are three critical values,
\begin{eqnarray}
h_{c1}=\mu\sqrt{\frac{\Delta^2_{\perp1}-t^2_{\perp}}{\Delta^2_{\perp1}-t^2_{\perp}+4\mu^2}},
\label{hc1}
\end{eqnarray}
\begin{eqnarray}
h_{c2}=\mu,
\label{hc2}
\end{eqnarray}
\begin{eqnarray}
h_{c3}&=&\sqrt{\Delta^2_{\perp1}-t^2_{\perp}+\mu^2}.
\label{hc3}
\end{eqnarray}
It is easy to see that, in general, $h_{c1}\leq h_{c2}\leq h_{c1}$.

In the case of $\mu\neq0$, from Eqs.(\ref{hc1}, \ref{hc2}) one can see that $h_{c1}\neq h_{c2}\neq 0$.
In this case, one possible type of Fermi surface topology with no effective Fermi surface for both $\textit{l}=\pm1$ occurs when $h<h_{c1}$. In this situation gap opens and pairing becomes of BCS-type (see \ref{fig1}a). The value of energy gap is 2$h_{c1}$. But for $\mu=0$, this regime vanishes and the pairing becomes gapless.

In the range  $h_{c1}<h<h_{c2}$ different type of Fermi surface topology takes place. For $\textit{l}=1$, there is no effective Fermi surface and excitation is gapful, whereas the branch $\textit{l}=-1$ crosses Fermi level in two points and consequently generates two effective Fermi surfaces, as depicted in Fig. \ref{fig1}b. This Fermi surface topology corresponds to Sarma pairing in the phase diagram and vanishes at $\mu=0$.

The other type of Fermi surface topology can be realized for $h_{c2}<h<h_{c3}$. In this range, there is no crossing point between branch $\textit{l}=1$ and Fermi surface, while the branch $\textit{l}=-1$ crosses Fermi surface in one point with a large Fermi momentum. Therefore, as is shown in Fig. \ref{fig1}c, because of large Fermi surface, the first excitation band goes to N state and the second band remains gapped. This topology defines new S phase which corresponds to the gapped mixed (GM) regime in the phase diagram.

The last possible Fermi surface topology can be determined when $h_{c3}<h$. In this range, both branches $\textit{l}=\pm1$ have one effective Fermi surface (see Fig. \ref{fig1}d). The branch $\textit{l}=-1(1)$ crosses Fermi surface with a large (small) Fermi momentum and goes to the N (S) state. It is interesting to note that splitting of Fermi surfaces by h in this case results in electron-like and holed-like Fermi surfaces in the absence of superconductivity. So, interlayer superconductivity, partially pairs n- and p-type momentum states. The corresponding phase will be referred to as pn phase \cite{HosseiniPN} in the phase diagram.

\begin{figure}[!tb]
\begin{center}
\includegraphics[width=8.4cm]{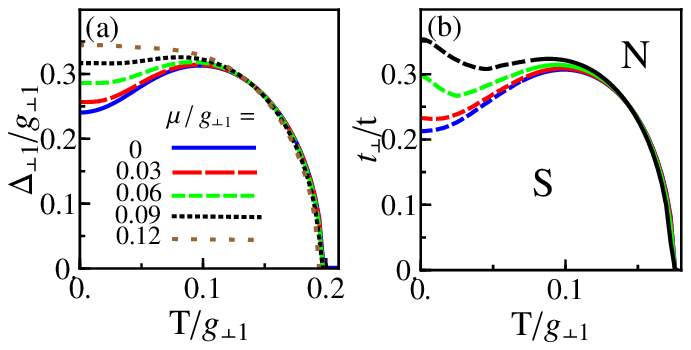}
\caption{(color online) For symmetric doping of bilayer graphene, $h=0$, (a) The plot of gap solution, $\Delta_{\perp1}$, versus $T$ for different levels of the mean doping, $\mu$, at $t_{\perp}=0$. (b) The mean field phase diagram of the chiral superconductivity showing dependence on the interlayer hopping energy, $t_{\perp}$, and the temperature, $T$, for $\mu/g_{\perp1}=0,0.03,0.06,0.09$ (from bottom to top, respectively). The line of transition from N to S phase is shown in which dashed and solid parts indicate the first and the second order transitions, respectively. }
\label{fig2}
\end{center}
\end{figure}
According to the above four different topologies for effective Fermi surfaces, the gap equation has different solutions.

\section{Interlayer superconductivity in doped bilayer graphene} \label{bilayer}
We obtain gap equation by minimizing the thermodynamic potential
\begin{eqnarray}\label{eq:Thermo}
\Omega_S=\frac{\Delta^2_{\perp1}}{g_{\perp1}}-\frac{1}{\beta}\sum_{\bf k,\gamma,\textit{l},\alpha}\ln(1+e^{-\beta \alpha E_{\bf k\textit{l}}^{\gamma}}),
\end{eqnarray}
with respect to the order parameters $\Delta_{\perp1}$, $\partial\Omega/\partial\Delta_{\perp1}=0$. By solving the gap equation self-consistently, one can obtain $\Delta_{\perp1}$.

For symmetric doping of the bilayer graphene, $h=0$, the solution of the gap equation, $\Delta_{\perp1}$ versus temperature, and the phase diagram of the chiral superconductivity in the $t_{\bot}-T$ plane, respectively, are shown in Fig.~\ref{fig2} for different levels of the mean doping. In Fig. \ref{fig2}a, the low temperature solutions of the gap equation increase with increasing the mean doping and at high mean doping, PA decreases monotonically with temperature resembling BCS-like behavior. This could be understood as follows: In the absence of doping, S state is gapless and the N fermions will exist as well and this weakens PA; in the presence of the mean doping, gap opens and PA strengthens. Finally, for high mean doping, Fermi surface becomes large and band structure is not important same as conventional s-wave symmetry, which results in BCS-like behavior of PA.
The phase diagram of the chiral superconductivity in
the $t_{\bot}-T$ plane is shown in Fig. \ref{fig2}b and the line of transition from N to S phases is shown which dashed and solid parts indicate the first- and the second-order phase transitions, respectively. One can see that low temperature region of S state increases by the mean doping. This is due to increase of PA by the mean doping [shown in Fig. \ref{fig2}a]. Since $t_{\bot}$ normally couples the two layers of the bilayer graphene, larger values of $t_{\bot}$ are required to vanish larger PA.
\par
\begin{figure}[!tb]
\begin{center}
\includegraphics[width=8.4cm]{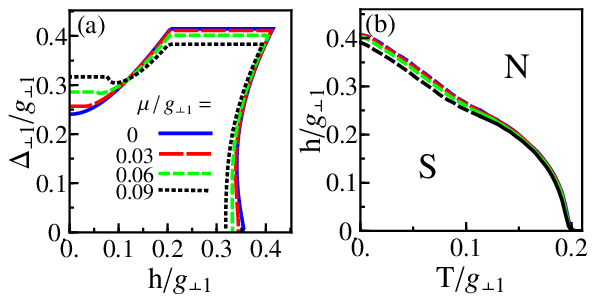}
\caption{(color online) For asymmetric doping of bilayer graphene, $h\neq0$, (a) The plot of none zero solutions of gap equation, $\Delta_{\perp1}$, versus $T$ for different levels of the mean doping $\mu$ at $t_{\perp}=0$. (b) The mean field phase diagram of the chiral superconductivity showing dependence on the doping difference, $h$, and the temperature, $T$, for $\mu/g_{\perp1}=0,0.03,0.06,0.09$ (from top to bottom, respectively). The line of transition from N to S phase is shown in which dashed and solid parts indicate the first and the second order transitions, respectively.}
\label{fig3}
\end{center}
\end{figure}
In the case of asymmetric doped bilayer graphene, $h\neq0$, the solution of the gap equation, $\Delta_{\perp1}$, versus the doping difference, $h$, at zero temperature and the phase diagram of the chiral superconductivity in $h-T$ plane are shown in Fig.~\ref{fig3} for different levels of the mean doping and $t_{\bot}=0$. As depicted in Fig. \ref{fig3}a, at finite mean doping the behavior of $\Delta_{\perp1}$ with respect to $h$ is constant up to the certain value of $h_{c1}$, decreases until develops a minimum at $h_{c2}$ and then increases, at a critical $h$, PA becomes constant and at $h_{c3}$ abruptly goes to zero. These four steps of behavior of $\Delta_{\perp1}$ correspond to the four different topologies of the Fermi surface as discussed above. Approaching $\mu=0$, BCS and Sarma states disappear completely. The phase diagram of (h-T) plane is shown in Fig. \ref{fig3}b, for different levels of the mean doping. The stable S region is separated from N region by phase transition line in which dashed and solid parts of the lines indicate the first and the second order phase transitions, respectively. At low (high) temperature and high (low) doping difference, the phase transition is the first (second) order. By increasing the mean doping, the range of the stable S state decreases.
\begin{figure}[!tb]
\begin{center}
\includegraphics[width=5.5cm]{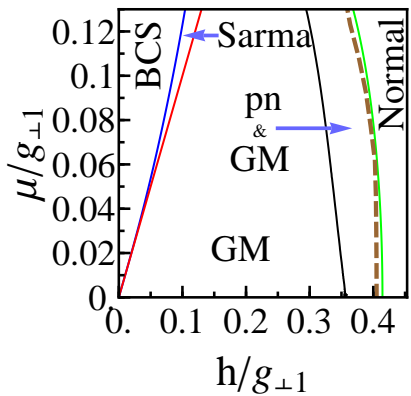}
\caption{(color online) For asymmetric doping of bilayer graphene, $h\neq0$, phase diagram of S states in $(\mu-h)$ plane is shown at zero temperature. Solid thin lines indicate boundaries of different phases and thick dashed line represents first order phase transition from S to N state.}
\label{fig4}
\end{center}
\end{figure}
\par
In Fig. \ref{fig4}, the phase diagram of S state in $(\mu-h)$ plane is shown at zero temperature. The solid thin lines indicate the boundaries of the different phases and the thick dashed line represents the first order phase transition from S to N state. One can see that BCS, Sarma and GM phases are stable, while pn phase is unstable. At $\mu=0$, BCS and Sarma states vanish and the only stable state is GM phase. Upon increasing $\mu$, the ranges of BCS and Sarma phases increase, with this property that the range of BCS state is larger than Sarma phase.
\par

\begin{figure}[!tb]
\begin{center}
\includegraphics[width=4.3cm]{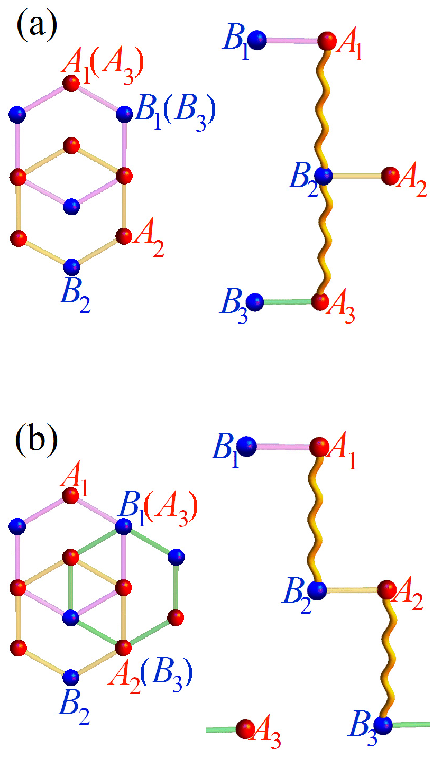}
\caption{(color online) Lattice structure of trilayer graphene. (a) Lattice structure of ABA-stacked trilayer graphene in which sublattices $A_1, B_2$ and $A_3$ from different layers are overlapped (left panel). In the right panel, interlayer correlations between electrons of sublattices $A_1$ and $B_2$ ($B_2$ and $A_3$) from first and second (second and third) layers are indicated by wavy lines. (b) Lattice structure of ABC-stacked trilayer graphene in which sublattices $A_1$ and $B_2$ from first and second layers and also $A_2$ and $B_3$ from second and third layers are overlapped (left panel). S correlations between interlayer nearest neighbor sublattices are shown in the right panel.}
\label{fig6}
\end{center}
\end{figure}
\section{Interlayer superconductivity in trilayer graphene} \label{trilayer}
In this section, we develop interlayer superconductivity to the case of trilayer graphene. Trilayer graphene consists of a monolayer graphene on the bottom of the bilayer graphene. There are two known manners of stacking in the bulk graphite which are called ABA and ABC \cite{TrilayerStaking} as shown in Fig. \ref{fig6}. In ABA stacking, the atoms of the bottommost monolayer lie exactly on the bottom of those of the top layer of bilayer, while for ABC stacking, one type of sublattice of the bottommost monolayer lies under the center of the hexagons in the bottom layer of the bilayer graphene. This subtle difference in stacking order remarkably affects on physical properties. Near the Dirac point, the electrons in ABC stacking behaves as massive fermions, while in ABA, electrons behave as both massive and massless Dirac fermions\cite{Trilayerband}. We consider the total Hamiltonian of $\pi$-electrons in the trilayer graphene without S correlation\cite{Trilayerband},
\begin{eqnarray}
H^{tri}_0=\sum^3_{\textit{l}=1}H^{mono}_{\textit{l}}+H^{ABA(C)}_{\perp},
\label{triHamiltonian}
\end{eqnarray}
where $H^{ABA(C)}_{\perp}$ is the nearest neighbor interlayer hopping Hamiltonian of ABA (ABC) stacked layers and has the following form,
\begin{eqnarray}
H^{ABA}_{\perp}=H^{AB}_{\perp}-t_{\perp}\sum_{\sigma, i}(b_{2,i,\sigma}^{\dagger}a_{3,i,\sigma}+H.c.),
\label{triABA}
\end{eqnarray}
\begin{eqnarray}
H^{ABC}_{\perp}=H^{AB}_{\perp}-t_{\perp}\sum_{\sigma, i}(a_{2,i,\sigma}^{\dagger}b_{3,i,\sigma}+H.c.),
\label{triABC}
\end{eqnarray}
 To induce the interlayer superconductivity, we introduce the nearest neighbor interlayer attractive interaction,
\begin{eqnarray}
V^{ABA}_{\perp}=V^{AB}_{\perp}-g_{\perp2}\sum_{\sigma,\sigma^{\prime},i}b_{2, i, \sigma}^{\dagger} b_{2, i, \sigma}a_{3, i, \sigma^{\prime}}^{\dagger}a_{3, i, \sigma^{\prime}},
\label{VpABA}
\end{eqnarray}
for ABA trilayer and
\begin{eqnarray}
V^{ABC}_{\perp}=V^{AB}_{\perp}-g_{\perp2}\sum_{\sigma,\sigma^{\prime},i}a_{2, i, \sigma}^{\dagger} a_{2, i, \sigma}b_{3, i, \sigma^{\prime}}^{\dagger}b_{3, i, \sigma^{\prime}},
\label{VpABC}
\end{eqnarray}
for ABC trilayer graphene. $g_{\perp2}$ couples the electrons of the sublattice $B_2$ ($A_2$) and $A_3$ ($B_3$) in ABA (ABC) arrangement.  The first term of Eqs.(\ref{VpABA}, \ref{VpABC}) can be decoupled by Eq.(\ref{Order}). Following discussion of Sec. \ref{s1}, we determine S order parameters between the second and the third layers
\begin{eqnarray}
\Delta_{i,\perp2}=-g_{\perp2}\langle b_{2, i, \downarrow}a_{3, i, \uparrow}+b_{2, i, \uparrow}a_{3, i, \downarrow} \rangle,
\label{OrderABA}
\end{eqnarray}
for decoupling the second term of Eq.(\ref{VpABA}) and
\begin{eqnarray}
\Delta_{i,\perp2}=-g_{\perp2}\langle a_{2, i, \downarrow}b_{3, i, \uparrow}+a_{2, i, \uparrow}b_{3, i, \downarrow} \rangle,
\label{OrderABC}
\end{eqnarray}
\begin{figure}[!tb]
\begin{center}
\includegraphics[width=5.5cm]{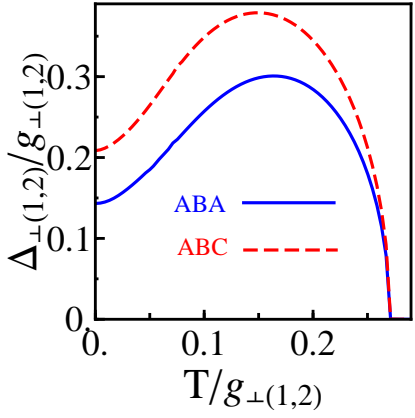}
\caption{(color online) Solutions of gap equations $\Delta_{\perp(1,2)}$ versus temperature for ABA and ABC stacking at $\mu=0$ and $t_{\perp}=0$. }
\label{fig7}
\end{center}
\end{figure}
for decoupling the second term of Eq. (\ref{VpABC}). In the momentum space, the total Hamiltonian of trilayer graphene ($H^{ABA(C)}=H_0^{tri}+V^{ABA(C)}_{\perp}$) can be diagonalized. With the procedure of Sec. \ref{bilayer}, we can calculate PA's for different stacking of trilayer. Fig. \ref{fig7} shows the nonzero stable solutions of $\Delta_{\perp1}$ and $\Delta_{\perp2}$ in terms of temperature for pure ABA and ABC trilayer at $t_{\perp}=0$. Same as the bilayer case, there is an enhancement of PA at the intermediate temperature. However, the values of ABC trilayer order parameter are larger than those of ABA trilayer, and as a result, in this case, simultaneous existence of $\Delta_{\perp1}$ and $\Delta_{\perp2}$ promote the interlayer superconductivity. we will discuss below the case of ABA stacking.
\begin{figure}[!tb]
\begin{center}
\includegraphics[width=8.5cm]{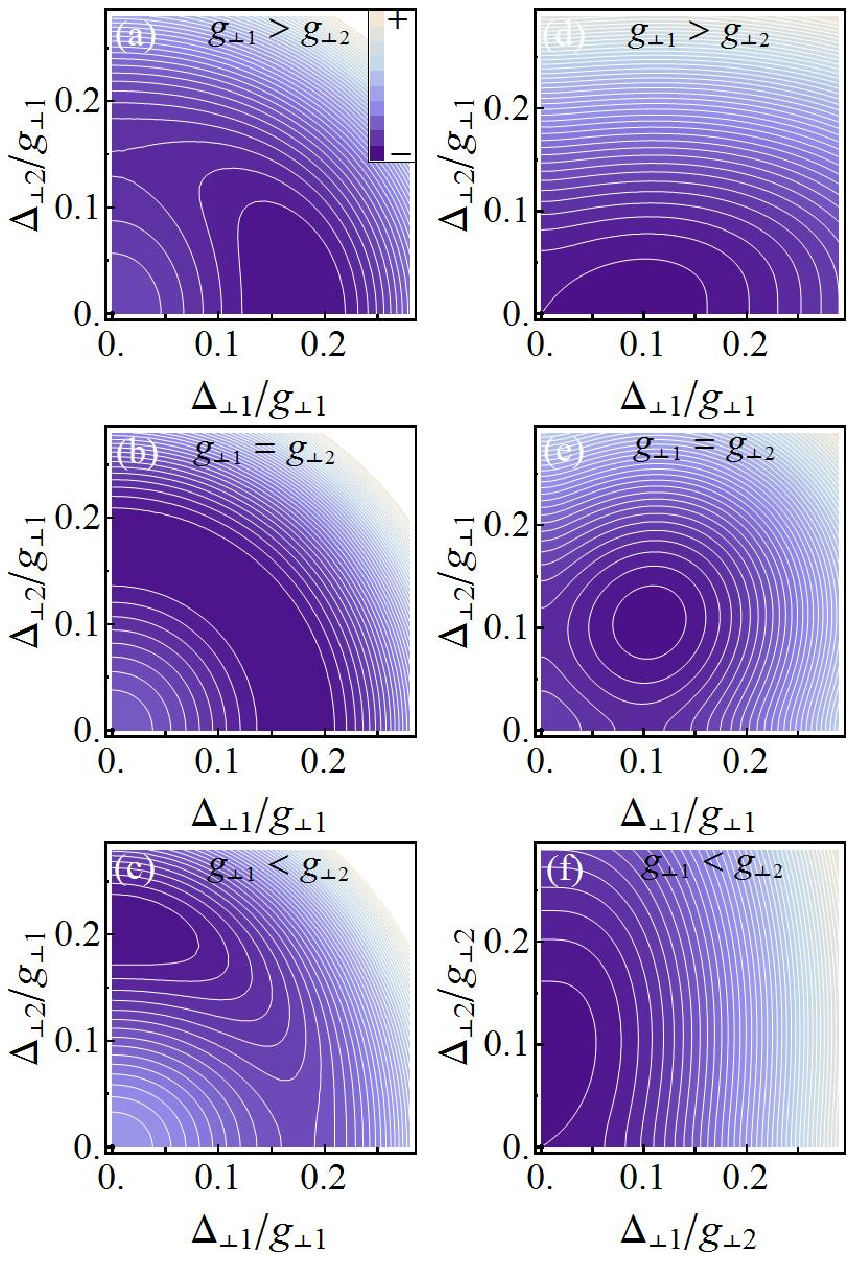}
\caption{(color online) Finite temperature thermodynamic potential contours in the plane of ($\Delta_{\perp1}$,$\Delta_{\perp2}$) showing stable points. In the case of ABA stacking, steps (a) $g_{\perp1}>g_{\perp2}$ ($g_{\perp1}/g_{\perp2}$=1.02), (b) $g_{\perp1}=g_{\perp2}$ , and (c) $g_{\perp1}<g_{\perp2}$ ($g_{\perp1}/g_{\perp2}$=0.98) indicate first order phase transition between $\Delta_{\perp1}$ and $\Delta_{\perp2}$, whereas in the case of ABC stacking steps (d) $g_{\perp1}>g_{\perp2}$ ($g_{\perp1}/g_{\perp2}$=3.11), (e) $g_{\perp1}=g_{\perp2}$ , and (f) $g_{\perp1}<g_{\perp2}$ ($g_{\perp1}/g_{\perp2}$=0.32) imply that second order phase transition occurs.}
\label{fig8}
\end{center}
\end{figure}
It is interesting to note that there is a mutual influence between the interlayer PA's. The manner of effect strongly depends on the type of layers arrangement. This effect can be investigated by studying of the competition between $\Delta_{\perp1}$ and $\Delta_{\perp2}$ with respect to the relative values of $g_{\perp1}$ and $g_{\perp2}$. In Fig. \ref{fig8}, the contour plot of the thermodynamic potential in the plane of ($\Delta_{\perp1}$ , $\Delta_{\perp2}$) for ABA- and ABC-stacked layers is shown for $\mu=0$ and $t_{\perp}=0$. In the ABA trilayer graphene, if $g_{\perp1}>g_{\perp2}$, the global (local) minimum of the thermodynamic potential lies on $\Delta_{\perp1}$ ($\Delta_{\perp2}$) axis (see Fig. \ref{fig8}a). By increasing $g_{\perp2}$ so that $g_{\perp1}=g_{\perp2}$, thermodynamic potential gives many stationary points where the
global minimum is a quarter of a circle shown in Fig. \ref{fig8}b. This behavior of global minimum is due to the symmetry of the parameter space.  By further increasing of $g_{\perp2}$, so that $g_{\perp1}<g_{\perp2}$, the minimum of the thermodynamic potential pass to the nonzero $\Delta_{\perp2}$ but zero $\Delta_{\perp1}$ (see Fig. \ref{fig8}c). On the other hand, by varying $g_{\perp1}$ and $g_{\perp2}$, stable point jumps between two axes implying first order phase transition between $\Delta_{\perp1}$ and $\Delta_{\perp2}$. The situation for ABC-trilayer is different. Fig. \ref{fig8}d shows that if $g_{\perp1}>g_{\perp2}$, the dominant PA is $\Delta_{\perp1}$ and the other has a small value. On the other hand, the thermodynamic potential has one point of global minimum without local minimum. This stationary point moves continually in the plane of ($\Delta_{\perp1}$ , $\Delta_{\perp2}$) by varying $g_{\perp1}$ and $g_{\perp2}$. At the same values of $g_{\perp1}$ and $g_{\perp2}$, PA's have equal magnitude (see Fig. \ref{fig8}e). Finally, $\Delta_{\perp2}$ will be larger than $\Delta_{\perp1}$ if $g_{\perp1}<g_{\perp2}$ as is shown in Fig. \ref{fig8}F. Therefore, changing of $g_{\perp}$'s exhibits the second order phase transition between the order parameters.

\section{conclusion} \label{con}

In conclusion, we have analyzed s-wave spin-triplet superconductivity in bilayer and trilayer graphene, which can be realized by an interlayer pairing between chiral electrons of different layers.  For bilayer graphene, we have shown that the gapless property of the interlayer S state of a pure sample and its associated temperature-induced enhancement of the S order parameter are suppressed by a finite mean doping.
 We have explored the possibility of realizing four distinct pairing regimes with different topologies of the Fermi surfaces by tuning the difference in the doping level of the two layer. These include the BCS regime in which two relevant branches of excitations are gapped and S (Fig. \ref{fig1} a), the Sarma regime with one branch gapped and the other gapless both S (Fig. \ref{fig1} b), gapped mixed regime with one branch gapped S and the other N (Fig. \ref{fig1} c), and pn regime with both branches being gapless, one S and the other N (Fig. \ref{fig1} d).

For trilayer graphene, we have shown that the interlayer pairing strongly depends on the type of the stacking, with ABC stacking supporting higher pairing amplitude than the ABA stacking. For an imbalanced  coupling strengths of the middle layer with the two outer layers, we distinct two phases which are characterized by the pairing between the middle layer with only one of the outer layers.   We have found a phase transition between these two phases by varying the ratio of the coupling strengths,  which is first order for the ABA stacking but second order for the ABC stacking.

\section{acknowledgement}
Authors gratefully acknowledge support by the Institute for Advanced Studies in Basic Sciences (IASBS) Research Council under grant No. G2012IASBS110. M. Z. thanks ICTP in Trieste for hospitality
and support during his visit to this institute.

\end{document}